\documentclass[preprint,showpacs,preprintnumbers,amsmath,amssymb]{revtex4}
\usepackage{appendix}
\usepackage{graphicx}

\begin{document}


\title{Faraday Rotation in a Disordered Medium}

\author{V. Gasparian$^{1}$ and Zh.S. Gevorkian$^{2,3,*}$}
\affiliation{
$^{1}$ California State University, Bakersfield\\
$^{2}$ Yerevan Physics Institute,Alikhanian Brothers St. 2,0036 Yerevan, Armenia. \\
$^{3}$ Institute of Radiophysics and Electronics, Ashtarak-2,0203, Armenia.\\
$^{*}$ gevork@yerphi.am}




\date{\today}

\begin{abstract}
The Faraday rotation angle $\Theta$ is calculated in a diffusive regime in a three dimensional disordered slab. It is shown that $\tan\Theta$ is (i) an oscillating function
of the magnetic field or the medium's internal properties, and (ii) proportional to the ratio of the inelastic mean free path $l_{in}$ to the mean free path $l$, that is to the average number of photon scatterings.
The maximum rotation is achieved at frequencies when the photon's elastic mean free path is minimal. We have obtained the rotation angle of polar backscattered light taking into account the maximally crossed diagrams.
The latter leads to an ellipticity in  the backscattered wave that can serve as precursor of weak
localization. The critical strength of magnetic field $B_c\sim g_c\sim \lambda/l$ beyond which rotation in backscattered wave disappears. The traversal time of an electromagnetic wave through the slab is estimated in a diffusive regime. The disorder enhanced the traversal time by an additional factor $ l_{in}/l$, in comparison with a free light propagation time. Comparison with the experimental data is carried out.
\end{abstract}



\maketitle

\section{Introduction.}
Since Anderson's pioneering work \cite{and}, where the concept of localization of an electron moving in a random potential was introduced, propagation of electromagnetic waves through disordered media have been attracting a lot of interest (see, for example, Ref. \cite{Neuw99} and references therein). One of the reasons for  such interest is the possibility of observing weak localization effects in disordered systems. Most of the papers were devoted to the backscattering peak where weak localization effects are manifest (see, e.g., Refs. \cite{ki84,al85,wm85,eta86,gl97} and for a recent experiment \cite{br12}). The influence of polarization effects on the backscattering peak was investigated in Ref. \cite{migc86}. Along with the transport properties of disordered media, magneto-optical effects in such systems capture broader attention. Particularly, in Ref. \cite{MacJon88}, the light backscattering peak in parity-non conserving disordered media was investigated. In such a medium rotation of the polarization plane is possible.

A Faraday rotation is a magneto-optical phenomenon that rotates the polarization of light. It occurs as an internal property of medium as well as under the external magnetic field. In both cases the dielectric permittivity tensor of the system becomes anisotropic. For example, in {\it GaAs} crystal large Faraday rotation \cite{pr03} origins because of the spin-orbit coupling. Faraday rotation is observed in astronomy. There  it is used for measurement of magnetic fields \cite{lm92}.

In the past decade many experimental papers \cite{tktifh06}-\cite{dwbwc11} have been devoted to the Faraday rotation in disordered media. In Refs. \cite{umfbi09,uch11,tlkrm11,dwbwc11} attention was drawn to the fact, that a large enhancement of the Faraday rotation can be obtained in nanomaterials incorporating several  nanoparticles and their composites. A garnet composite film (mostly a Bi-substituted yttrium iron garnet (Bi:YIG)) incorporating gold nanoparticles or a nanoparticles solution with gold coated $Fe_2O_3$ \cite{umfbi09,uch11,tlkrm11,dwbwc11} show a change in the sign of the Faraday rotation at the wavelengths corresponding to the surface plasmon absorption peak of the gold nanoparticles. Around the local surface plasmon resonance wavelengths a resonant transmittance attenuation and a large resonant enhancement of the Faraday rotation with a narrow bandwidth was obtained in Ref \cite{umfbi09}. Recently a large Faraday rotation was observed \cite{nature11} in a ultrathin graphene film.  The analogy with the Hall effect is revealed.

The standard effective medium approximations (Maxwell-Garnet and Bruggeman theories) \cite{choy99} describe the macroscopic properties of a medium using the average dielectric constant determined by  the relative fractions of its components. These theories show only qualitative agreement with the experimental data of Refs.\cite{umfbi09,uch11,tlkrm11,dwbwc11}. Many of the questions that have been raised during the experiments are still waiting for answers. For example, no corresponding analytical calculation has been available so far for a discrepancy in the bandwidth between the transmission attenuation and the resonant Faraday rotation (Ref. \cite{umfbi09}). Of particular importance is the influence of the disorder, induced by the presence of the impurities, on the Faraday angle and on several related quantities.

In the present paper we develop a theoretical approach for investigating the Faraday rotation for transmitted and  backscattered waves in three-dimensional disordered media. We show that the maximum Faraday rotation angle of the transmitted wave is achieved at the frequencies where the photon elastic mean free path is minimal.  In the backscattered wave the maximally crossed diagrams contribution to the Faraday rotation angle is very important. It leads to an ellipticity of the backscattered wave in the localized regime.

The paper is organized as follows.  In Sec. II we formulate the problem. The coherent potential approximation is discussed in Sec. III and appropriate contribution to the Faraday angle is calculated. In Sec. IV we present result for the diffusional contribution to the Faraday rotational angle. Diffusional and maximally crossed diagrams contributions are taken into account for Kerr effect in Sec. V.  Obtained analytical results are compared with experimental data in Sec. VI. Finally, the main conclusions are summarized in Sec. VIII. The Paper contains two  Appendices, which present some technical details useful for the
understanding of diffusional and maximally crossed diagrams contributions to the Kerr effect.
\section{Formulation of the Problem}
Let us consider the incidence of a linearly polarized electromagnetic wave on the optically active medium, see Fig.1.
\begin{figure}
\includegraphics[width=9cm]{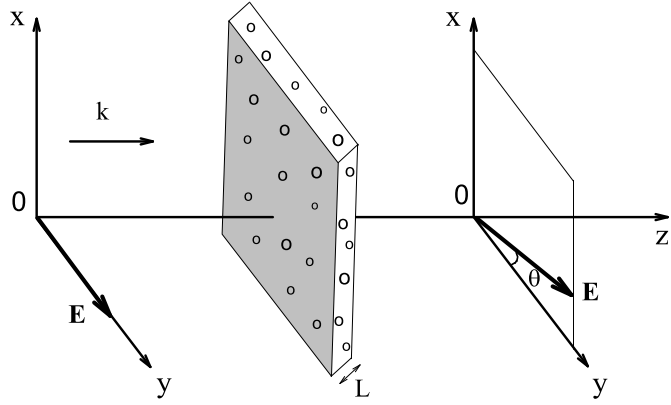}
\caption{Geometry of the problem. Incident wave is polarized on $0y$. After transmission through a slab the polarization vector rotates.}
\label{fig.1}
\end{figure}
We assume that the incident field is polarized on $0y$, the plane of incidence is $xz$, wave vector is directed on $0z$.
Because of the scattering transmitted field contains all the directions of wave vector. In order to determine the Faraday angle one should separate the coherent part that is the one directed on $0z$, see Fig.1. Faraday rotation angle is determined as follows
\begin{equation}
\tan\Theta=<\frac{E_x}{E_y}>,
\label{angle}
\end{equation}
where $E_{x,y}$ are the components of transmitted coherent wave and $<...>$ means averaging over the realizations of randomness. The phases of electric field components are strongly fluctuating in contrary to intensities. Therefore for a weakly scattering medium one can approximately take
\begin{equation}
\tan\Theta\approx\frac{<E_xE_y^*>}{<E_yE_y^*>}.
\label{inten}
\end{equation}
Maxwell equation for the electric field has the form
\begin{equation}
\nabla^2E_i(\vec r)+\varepsilon_{ij}(\vec r)\frac{\omega^2}{c^2}E_j(\vec r)=j_i(\vec r),\quad \bf \nabla\bf E=0
\label{maxw}
\end{equation}
where $\vec j$ is the source term and the dielectric tensor is determined as follows
\begin{equation}
\varepsilon_{ij}(\vec r)=(1+\varepsilon(\vec r))\delta_{ij}-ie_{ijk}g_k.
\label{ditensor}
\end{equation}
Here $e_{ijk}$ is the antisymmetric tensor, $\bf g$ is the gyration vector directed on the magnetic field direction when external field is applied. In the opticaly active medium without external field $\vec g=f\vec q$ and is parallel to the direction of photon propagation. $\varepsilon(\vec r)$ is the random part which is assumed Gaussian distributed with $\delta$-correlation function
\begin{equation}
<\varepsilon(\vec r)\varepsilon(\vec r^{\prime})>=\gamma\delta(\vec r-\vec r^{\prime}).
\label{dicorr}
\end{equation}
In order to carry out averaging over the randomness it is convenient to express all the quantities through the Green's functions
\begin{equation}
\nabla^2G^R_{ij}(\vec r,\vec r^{\prime})-\frac{\partial^2G^R_{kj}(\vec r,\vec r^{\prime})}{\partial r_i\partial_k}+\varepsilon_{ik}(\vec r)\frac{\omega^2}{c^2}G^R_{kj}(\vec r)=\delta_{ij}\delta(\vec r-\vec r^{\prime}).
\label{green}
\end{equation}
As was mentioned, the Faraday rotation angle in Eqs. (\ref{angle}) and (\ref{inten})  is determined by the coherent intensity. Hence, to separate the coherent intensities let introduce the intensity tensor at observation point $\vec R$ and in the direction of unit vector $\vec s$ following \cite{migc86}
\begin{equation}
I_{ij}(\vec R,\vec s)=2\int_0^{\infty}q^2dq\int d\vec r e^{-iq\vec s\vec r}<E_i(\vec R+\vec r/2)E_j^*(\vec R-\vec r/2)>.
\label{inttens}
\end{equation}
Using Eq.(\ref{inttens}) the Faraday angle, Eq.(\ref{inten}), can be represented in terms of a intensity tensor $I_{ij}(\vec R,\vec s)$
\begin{equation}
\tan\Theta\approx\frac{I_{xy}(L, \hat{\bf s_z})}{I_{yy}(L, \hat{\bf s_z})}.
\label{angten}
\end{equation}
Here $\hat{\bf s_z}$ is the unit vector on $z$ and $L$ is the thickness of the slab. Because of the symmetry the intensities depend only on the $z$ coordinate.

\section{ Coherent Potential Approximation}
Averaged Green's functions take into account the fact that the right-hand and the left-hand circular polarized lights see different refractive indexes in the medium \cite{MacJon88}
\begin{equation}
G_{ij}^{R,A}(\vec q)=\sum_{\alpha=\pm 1}\frac{\frac{1}{2}(\delta_{ij}-\hat{q_i}\hat{q_j}\mp i\alpha e_{ijm}\hat{q_m})}{(1-\alpha \hat{\bf q} \vec{\bf g}) k_0^2-q^2\pm ik_0/l}
\label{avgreen}
\end{equation}
where $k_0=\omega/c$, $\hat{\bf q}$ is a unit vector on $\vec q$, $l=6\pi/\gamma$ is the photon elastic mean free path. The superscripts R and A denote retarded and advanced Green's functions, respectively.

Faraday angle consists of two main contributions. One comes from the coherent potential approximation and another from the diffusional scattering of photon. Coherent potential contribution reads
 \begin{equation}
 \tan\Theta_p=\frac{\bar{E_x}(L)}{\bar{E_y}(L)},
 \label{cohap}
 \end{equation}
 where
\begin{equation}
\bar{E_i}(\vec R)=\int d\vec R^{\prime}<G_{im}^R(\vec R,\vec R^{\prime})> j_m(\vec R^{\prime}).
\label{bare}
\end{equation}
The source term $\vec j(\vec r)$ is chosen in such a way that the coherent part of transmitted electric field for $g=0$ be equal
\begin{equation}
\bar{E_i}(\vec r)=\delta_{iy}e^{ik_0z-z/2l}.
\label{chpart}
\end{equation}
Using Eqs.(\ref{avgreen}) and (\ref{chpart}), provided that $k_0l\gg 1$, one finds
\begin{equation}
j_i(\vec q)=2k_0(2\pi)^2i\delta_{iy}\delta(q_x)\delta(q_y).
\label{source}
\end{equation}
Substituting this expression into Eq.(\ref{bare}) and Eq.(\ref{cohap}) in the limit $g\ll 1$ and $k_0l\gg 1$, we obtain
 \begin{equation}
 \tan\Theta_p=e^{ik_0L}\left[\sin\frac{gk_0L}{2}-\frac{ig}{2}\cos\frac{gk_0L}{2}\right],
 \label{cohan}
 \end{equation}
where $g\equiv g_z$. It follows from Eq.(\ref{cohan}) that coherent potential approximation contribution, under the conditions $g\ll 1$ and $k_0l\gg 1$, becomes inessential because contains strongly oscillating term $\exp(ik_0L)$  when $k_0L\gg 1$. Hence, any averaging over the wavelength of incident photon will make this contribution negligible. Therefore below we concentrate our attention on the diffusional contribution to the Faraday rotation angle.

\section{Diffusional Contribution}
Using Eq.(\ref{inttens}), we have
\begin{eqnarray}
&I_{ij}^D(\vec R,\vec s)=2\int_0^{\infty}q^2dq\int_{\vec R^{\prime}, \vec r^{\prime}, \vec r,\vec \rho_1..\vec \rho_4}e^{-iq\vec s\vec r}G^R_{im}(\vec R+\vec r/2,\vec \rho_1)G_{nj}^A(\vec \rho_3,\vec R-\vec r/2)\times \nonumber \\
&P_{mnhs}(\vec \rho_1,\vec \rho_2,\vec \rho_3,\vec\rho_4)G^R_{hf}(\vec\rho_2,\vec R^{\prime}+\vec r^{\prime}/2)G^A_{gs}(\vec R^{\prime}-\vec r^{\prime}/2,\rho_4)j_f(\vec R^{\prime}+\vec r^{\prime}/2)j_g(\vec R^{\prime}-\vec r^{\prime}/2),
\label{difcon}
\end{eqnarray}
where $P_{mnhs}(\rho_1,\rho_2,\rho_3,\rho_4)$ is the sum of ladder diagrams
\begin{figure}
\includegraphics[width=9cm]{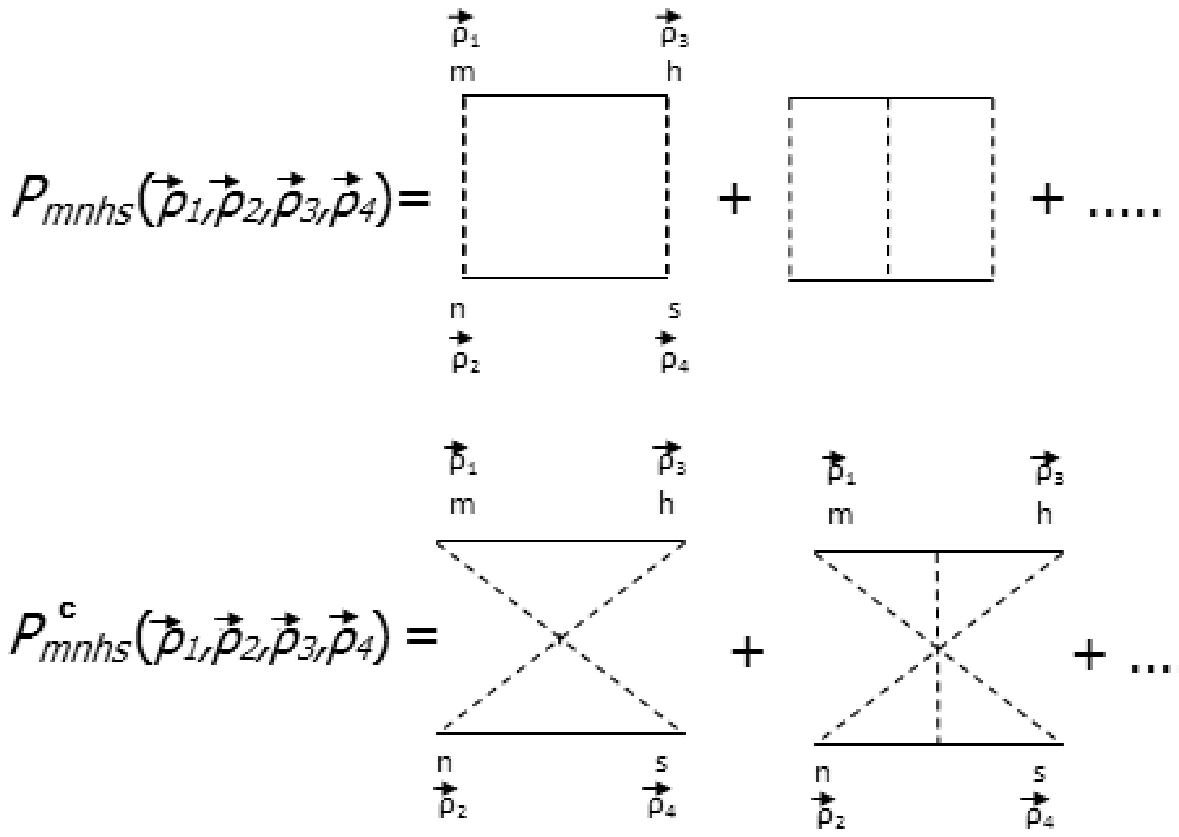}
\caption{Dashed line is the correlation function of random field Eq.(\ref{dicorr}) and the solid lines are the photon retarded and advanced averaged Green's functions Eq.(\ref{avgreen}). }
\label{fig.2}
\end{figure}
which was found in innumerable papers, see for example, \cite{migc86}. We will calculate the intensity tensor in the limit $g\to 0$. Therefore, it is enough to calculate the diffusion propagator for $g=0$. The main contribution to the integrals in Eq.(\ref{difcon}) gives the term containing the diffusion pole. The corresponding mode has the form (see for example, Ref. \cite{migc86})
\begin{equation}
 P_{mnhs}(\vec\rho_1,\vec\rho_2,\vec\rho_3,\vec\rho_4)=\delta_{mn}\delta_{hs}\delta(\vec \rho_1-\vec \rho_2)\delta(\vec \rho_3-\vec \rho_4)P(\vec \rho_1-\vec\rho_3),
 \label{difmod}
 \end{equation}
 where $P(\vec\rho)=\int\frac{d\vec K}{(2\pi)^3}P(\vec K)e^{i\vec K\vec\rho}$ and
\begin{equation}
P(\vec K)=\frac{3\gamma}{l/l_{in}+K^2l^2}.
\label{difprop}
\end{equation}
Here $l_{in}$ is the inelastic mean free path of a photon in the medium and it is assumed that $\lambda\ll l\ll l_{in},L$.

Fourier transforming Eq.(\ref{difcon}), substituting the expression $P(\vec K)$ into it, and using  Eq.(\ref{bare}), one obtains
\begin{equation}
I_{xy}^D(\vec R,\hat{\bf s_z})=12\gamma\l_{in}\int_0^{\infty}q^2dq\int_{-\infty}^{\infty}\frac{dK_z}{2\pi}\frac{e^{iK_zZ}G_{xm}^R(K_z/2+q)G_{my}^A
(K_z/2-q)}{(1+K_z^2l^2)(1+K_z^2ll_{in})}
\label{intens}
\end{equation}
where
\begin{equation}
G^{R,A}_{xm}(K_z/2+q)=\sum_{\alpha=\pm 1}\frac{\frac{1}{2}(\delta_{xm}\mp i\alpha e_{xmz})}{(1-\alpha g_z)k_0^2-(K_z/2+q)^2\pm ik_0/l}.
\label{greenz}
\end{equation}
The integral over $K_z$ can be calculated exactly closing the integral contour in the complex $K_z$ plane in the upper half ($Z>0$). Note that the contributions from the poles $K_z=i/l$ and $K_z=i\sqrt{1/ll_{in}}$ in the sum over polarizations cancel each other in the limit $g\to 0$ and $k_0l\gg 1$. Substituting Eq.(\ref{greenz}) into Eq.(\ref{intens}) and subsequently calculating the integrals over $K_z$ and $q$, we obtain
\begin{equation}
I_{xy}^D(\vec R,\hat{\bf s_z})=6\pi^2\frac{l_{in}}{l}e^{-Z/l}\sin gk_0Z.
\label{xycomp}
\end{equation}
In deriving this expression we assume that $gk_0l\ll 1$,  however $k_0l\gg 1$.
The quantity $I_{yy}^D(\vec R,\hat{\bf s_z})$ can be calculated in the same way as $I_{xy}^D(\vec R,\hat{\bf s_z})$. In this case the $g=0$ approximation can be
used, because  $I_{xy}^D(\vec R,\hat{\bf s_z})$ is already proportional to $g$. Substituting Eq.(\ref{bare}) into Eq.(\ref{intens}), one has
\begin{equation}
I_{yy}(\vec R,\hat{\bf s_z})=4\pi^3e^{-Z/l}.
\label{yycomp}
\end{equation}
Using the above calculated expressions for $I_{xy}^D(\vec R,\hat{\bf s_z})$, $I_{yy}^D(\vec R,\hat{\bf s_z})$ and Eq.(\ref{angten}), we finally obtain the desired result for the Faraday rotation angle with the diffusional contribution
\begin{equation}
\tan\Theta^D\approx\frac{3}{2\pi}\frac{l_{in}}{l}\sin gk_0L.
\label{final}
\end{equation}
Equation (\ref{final}) with Eq. (\ref{diffcross}) (see below) represent the
central results of this work. $\tan\Theta^D$ presents the asymptotically exact
($g \to 0$, $k_0l\gg 1$) expression for the Faraday rotation in 3d disordered systems.
Comparing the two formulae, $\tan\Theta^p$ and $\tan\Theta^D$, we see that the latter oscillates less, due to the fact that $g\to 0$ in the argument of the $sin$ function. Therefore, the diffusion contribution to the Faraday rotation, Eq.(\ref{final}), will dominate compared to the coherent potential approximation contribution, Eq.(\ref{cohan}). This is valid provided that diffusional scattering is realized in the disordered system.

We remind that Eq.(\ref{final}) has been derived assuming that $gL/l\ll 1$ however $gL/\lambda$ is not necessarily small.
It is worth noticing that the ratio $l_{in}/l$ is the average number of scatterings  of the photon in the medium. If the thickness $L$ of a slab is thinner than the length $\sqrt{ll_{in}}$ then the diffusional trajectories are cut on the system size and the above mentioned number of scatterings becomes equal to $L^2/l^2$ (see Refs. \cite{And85,Gev98}). In this particular case, Eq.(\ref{final}) can be rewritten as
\begin{equation}
\tan\Theta^D\approx \frac{3}{2\pi T^2}\sin gk_0L,
\label{final1}
\end{equation}
taking into account that in the diffusional regime the transmission coefficient $T$ of a disordered slab is of order $l/L$.
It follows from Eqs.(\ref{final}) and (\ref{final1}) that the maximum of the Faraday angle is achieved at the frequencies where the photon elastic mean free path $l$ is minimal. In the case of Eq. (\ref{final1}) ($L<\sqrt{ll_{in}}$), it is obvious that the maximum of the Faraday rotation corresponds to the minimum of the transmission coefficient and in its proximity. This type of  dependence was confirmed experimentally in Refs. \cite{umfbi09,uch11}.
However the Faraday rotation peak will be steeper because of stronger dependence on $l(\omega)$ (see below for more details). Another pecularity of our result, Eq.(\ref{final}), is that it implies an oscillating dependence of the rotation on $g$, that is on a magnetic field, or the medium's internal properties. Note that the non-monotonous dependence of the Faraday rotation on the magnetic field is observed in Ref.\cite{tlkrm11}.
\section{Rotation in the backscattered light: Kerr effect}
For completeness, we now turn to the discussion of the Faraday rotation angle for the backscattered wave using the procedure outlined above. The main difference now is that one should take into account the contribution of the maximally-crossed diagrams (Ref. \cite{ln65}) in Fig.2, which goes beyond the ladder diagram approximation, (i.e., regular diffusion).
Another important point to mention also is that in the Kerr effect the boundary conditions at $z=0$ are important.

For a brief sketch of the derivation of the backscattering peak note, that the diffusion propagator $P$ should fulfill the boundary condition at $z=0$. Therefore, instead of the former $P_{mnhs}$, we, based on the method of images, use a modified
$P(\vec \rho, z_1,z_2)=P(\vec \rho, z_1-z_2)-P(\vec \rho, z_1+z_2)$. Here $\vec\rho$ is a two dimensional vector in the $xy$ plane and we assume that the following
inequalities are met: $g\ll 1$, $k_0l\gg 1$ and $gk_0l\ll 1.$ The details of the calculations of the backscattering rotation is given in Appendices A and B. The final result ($l_{in}=\infty$, i.e. no absorption) reads
\begin{equation}
\tan\Theta_b^D=-\frac{3}{2}gk_0l\quad \tan\Theta_b^C=\frac{ig}{4},
\label{diffcross}
\end{equation}
where $\Theta_b^{D,C}$ are the diffusional and maximally crossed diagrams contribution to the Faraday rotation angle of the backscattered light, respectively. Note that maximally-crossed diagrams leads to a pure imaginary Faraday angle which indicates the onset of ellipticity for the backscattered wave. As is seen from Eq.(\ref{diffcross}) the diffusional contribution is dominant in the diffusion regime $k_0l\gg 1$ in contrary to the maximally crossed diagrams contribution which becomes relatively large in the localized regime $k_0l\sim 1$. In this regime, $k_0l\sim 1$, the diffusional and maximally crossed diagrams contributions are of the same order and maximum of rotation angle is achieved. Physically, such a behavior can be understood within the existing relation between the Faraday rotation angle and the time spent by the light in a medium \cite{gas,gas1} (see below). Indeed, in the diffusive regime the backscattering time is larger for low scatterer concentration (or for larger photon mean free path). With increasing the scatterer concentration (the photon mean free path becomes smaller) the backscattering time increases because of the localization effects (or because of the ellipticity of the polarization of the reflected wave). This transparent picture serves as a basis for qualitative understanding of asymptotic behavior of backscattered wave in two regimes. Thus, the onset of ellipticity in the backscattered wave can serve as a precursor of light weak localization.

Note, that in the limit $gk_0l\gg 1$ the diffusional and maximally crossed diagrams contributions to the rotation angle of backscattering wave are very small (see Eqs. (\ref{idfinal}) and (\ref{maxcontr})). Hence for the backscattered wave the maximal rotation is achieved for $g \sim 1/k_0l$.

In concluding this section, let us estimate the traversal time of an electromagnetic wave in 3d disordered media. According to Ref. \cite{gas} the interaction time in a slab is closely related to the Faraday rotation. Hence, one might expect, that the traversal time, $\tau$,  of an electromagnetic wave through a 3d disordered slab will be larger compared with the time, $\tau_0=Ln_0/c$, spent in a dielectric with open boundaries. Following Ref. \cite{gas} and expanding Eq. (\ref{final}) in powers of $g$, and keeping linear terms only we get an estimated value for time  $\tau\approx(l_{in}/l)\tau_0.$ Thus, the disorder enhanced the traversal time $\tau$ by an additional factor $l_{in}/l$, in comparison with a free light propagation time $\tau_0$.
\section{Comparison with the experiment}
Most of the experiments on Faraday rotation in disordered media \cite{tktifh06}-\cite{tlkrm11} are carried out with the films of thicknesses much less than the light wavelengths. Our approach cannot be directly applied to these systems because the randomness in this case is nearly two-dimensional contrary to our $3d$ randomness. We will compare  our results with the experimental data of the work  Ref. \cite{dwbwc11}, where the sample has $3d$ randomness. First let us estimate the dimensionless constant $g$ for a water noting that in Ref. \cite{dwbwc11} nanoparticles are randomly embedded in the water. It follows from (\ref{greenz}) that $g=(\varepsilon_{+}-\varepsilon_{-})/2=n(n_{+}-n_{-})$ and $n=(n_{+}+n_{-})/2$. $\varepsilon_{\pm}$ and $n_{\pm}$ are dielectric constants and refraction indexes of right-hand and left-hand circularly polarized light at $632nm$ wavelength, respectively. The difference $\Delta n=n_{+}-n_{-}$ is related to the Verdet constant $\nu$: $\theta=\nu B L=\pi\Delta nL/\lambda$. Taking $B=1T$,  one has $\Delta n=\lambda\nu/\pi$. For a water at $\lambda=632nm$,$n=1.33$ and $\nu=3.4rad/Tm$, see for example, \cite{verdet}. Therefore $g=n\Delta n\sim 10^{-6}$. We see that for a water at $B=1T$, $g$ is extremely small and the limit $g \to 0$ in our theory is justified. The elastic photon mean free path $l$ in a system of randomly embedded in a dielectric host medium of metallic spheres can be estimated as $l={8\pi c^4}/{(9vf\omega^4)}$,
where $v$ is the volume of a single sphere and $f$ is the volume fraction of spheres in the medium \cite{john85}. We take $v\approx 4\pi a^3/3$, where $a\approx 85nm$ is the average size of particles coated with gold in an aggregate state and $f=0.035$ \cite{dwbwc11}. Substituting these values and $\lambda=632nm$ into expression of $l$, one gets $l\approx 5\lambda$. Now let us estimate the inelastic mean free path $l_{in}$, which is defined as $l_{in}={\lambda}/{(2\pi Im\varepsilon_{eff})}$ \cite{john85,asb86} ($Im\varepsilon_{eff}={9f\varepsilon_2}/{(\varepsilon_1-1)^2}$ is the imaginary part of the Maxwell-Garnett effective dielectric constant and $\varepsilon_1$, $\varepsilon_2$ are the real and imaginary parts of the dielectric constant of gold at incident wavelength $\lambda=632nm$, respectively). Taking
$\varepsilon_1\approx -10.7$ and $\varepsilon_2\approx 3.23$, Ref.\cite{jocher72}, one has $l_{in}\approx 21.4\lambda$ and it is easy to convince oneself that the conditions for photon diffusion $\lambda\ll l(\lambda)\ll l_{in}(\lambda)\ll L$ are realized in the experiment \cite{dwbwc11}. In this case the additional large multiplier $l_{in}/l\sim 4$ arises in the Verdet constant calculated within the effective Maxwell-Garnett theory. This additional factor could explain the discrepancy between the theoretical calculations and the experimental data, mentioned in Ref. \cite{dwbwc11}. Multiplying the Verdet constant of water $3.4$ by $l_{in}/l\sim 4$ and subtracting the water value, for the Verdet constant of solution, one gets  $\nu\sim 10.2$ which is in a very good agreement with the corresponding experimental value \cite{dwbwc11}.
\section{Summary.} We have investigated the Faraday rotation in a disordered medium. The rotation angle is calculated to leading order of parameters $gk_0L$, $1/k_0l$. It is shown that in the diffusive regime Faraday rotation is proportional to the ratio of the inelastic mean free path $l_{in}$ to the mean free path $l$, that is to the average number of photon scatterings in the medium. For the Faraday angle the magnetic field oscillating dependence is predicted.
If the thickness $L$ of a slab is thinner than the length $\sqrt{ll_{in}}$ then the Faraday rotation is inverse proportional to the square of the transmission coefficient. The maximum rotation is achieved at frequencies where the photon elastic mean free path is minimal. We have calculated the Faraday rotation of backscattered waves. Maximally-crossed diagrams lead to an ellipticity of backscattered wave. In the localized regime $k_0l\sim 1$ the diffusional and maximally crossed diagrams contributions Eq.(\ref{diffcross}) to the rotation angle are of the same order. Therefore appearing of ellipticity in the backscattered wave can serve as a precursor of a weak localization of the light. In the backscattered wave maximal rotation is achieved for $g\sim 1/k_0l$. Beyond this critical value rotation angle rapidly disappears. The traversal time of an electromagnetic wave through the slab is estimated in a diffusive regime. Comparison with the experimental data is carried out. To summarize, although the main result for Faraday rotation, Eq. (\ref{final}), is approximate, it nevertheless predicts correctly  many of the peculiar features of most  experimental results \cite{tktifh06}-\cite{dwbwc11}, discussed in this paper.

{\bf Acknowledgments.}
We are grateful to T. Meyer, A. Akopian and O. del Barco for helpful comments and discussions. V.G. acknowledges partial support by FEDER and the Spanish DGI under project no. FIS2010-16430.

\setcounter{section}{0}
\renewcommand\thesection{\setcounter{equation}{0}Appendix \Alph{section}:}
\renewcommand\theequation{\Alph{section}\arabic{equation}}
\section[\hspace{2.4cm}Diffusional contribution to the Kerr effect]
{Diffusional contribution to the Kerr effect}
We derive first the diffusional contribution to the Kerr effect. i.e., $\tan\Theta_b^D=-\frac{3}{2}gk_0l$ of Eq. (\ref{diffcross}). To do this, let us note that the diffusional contribution to the intensity tensor, see Fig.2, takes the form
\begin{equation}
I_{ij}^D(\vec R,\vec s)=2\int_0^{\infty}q^2dq\int d\vec re^{-iq\vec s\vec r}\int d\vec r_1 d\vec r_2 G^R_{im}(\vec R+\vec r/2,\vec r_1)G_{mj}^A(\vec r_1,\vec R-\vec r/2) P(\vec r_1,\vec r_2)\bar E_h(\vec r_2)\bar E_h^*(\vec r_2),
\label{ker}
\end{equation}
where
\begin{equation}
P(\vec r_1,\vec r_2)=P(\vec\rho,z_1-z_2)-P(\vec\rho,z_1+z_2).
\label{image}
\end{equation}
Here $\vec\rho$ is the projection of $\vec r_1-\vec r_2$ in the $xy$ plane. $P(\vec r)$ is determined by three dimensional Fourier transform of Eq.(\ref{difprop}) and the coherent field $\bar E_h(\vec r_2)$ is determined by Eq.(\ref{chpart}).
Going through Fourier transforms in Eq.(\ref{ker}), one gets
\begin{equation}
I_{ij}^D(\vec R,\vec s)=2 \int_0^{\infty}q^2dq\int\frac{dq_{1z}dK_z}{2\pi}\delta(q_{1z}-K_z/2-qs_z)G^R_{im}(q_{1z})G^A_{mj}(K_z-q_{1z})e^{iK_zZ}I(K_z),
\label{fur}
\end{equation}
where
\begin{equation}
I(K_z)=\int_0^{\infty}dz_1dz_2P(0,z_1,z_2)e^{-iK_zz_1-z_2/l}
\label{ikz}
\end{equation}
and
\begin{equation}
P(0,z_1,z_2)=\int d\vec\rho P(\rho,z_1,z_2).
\label{pzet}
\end{equation}
Here we assume that $s_{\rho}=0$. Substituting Eqs.(\ref{difprop}) and (\ref{image}) into Eq.(\ref{pzet}), we obtain
\begin{equation}
P(0,z_1,z_2)=\frac{3\gamma}{2l^2}\left[z_1+z_2-|z_1-z_2|\right].
\label{furimage}
\end{equation}
Note that the expression (\ref{furimage}) was derived ignoring the absorption and assuming that $l_{in}=\infty$.
Substituting Eq.(\ref{furimage}) into Eq.(\ref{ikz}), one has
\begin{equation}
I(K_z)=\frac{3\gamma i}{K_z}(e^{-iK_zL}-1)+\frac{3\gamma i}{K_z-i/l}.
\label{finikz}
\end{equation}
Note, that in order to arrive to Eq. (\ref{finikz}), we cut the diverging integral on the upper limit at the system thickness $L$. For the backscattered direction $s_z=-1$ and for $\vec R=0$ the intensity $I_{xy}^D(\vec R,\vec s)$, Eq.(\ref{fur}), can be written in the form
\begin{equation}
I_{xy}^D(0,-1)=2 \int_0^{\infty}q^2dq\int\frac{dK_z}{2\pi}G^R_{xm}(K_z/2-q)G^A_{my}(K_z/2+q)I(K_z).
\label{intenback}
\end{equation}
Substituting the Green's functions, Eq.(\ref{avgreen}), into Eq.(\ref{intenback}) and keeping only linear terms on $K_z$, one has
\begin{equation}
I_{xy}^D(0,-1)=I(g)-I(-g),
\label{ige}
\end{equation}
where
\begin{equation}
I(g)=i\int_0^{\infty}dq\int\frac{d K_z}{2\pi}\frac{I(K_z)}{(K_z+\frac{(1-g)k_0^2-q^2}{q}+\frac{ik_0}{ql})(K_z+\frac{q^2-(1+g)k_0^2}{q}+\frac{ik_0}{ql})}.
\label{iplusge}
\end{equation}
We will integrate Eq.(\ref{iplusge}) over $K_z$ closing integral contour in the bottom half of complex plane where the function $I(K_z)$ has no poles. The contributions to the integral come from the poles $K_z=-\frac{(1-g)k_0^2-q^2}{q}-\frac{ik_0}{ql}$ and $K_z=-\frac{q^2-(1-g)k_0^2}{q}-\frac{ik_0}{ql}$. Using Eqs.(\ref{finikz}) and (\ref{ige}) and taking integral over $K_z$ in the limit $g\to 0$, we find
\begin{eqnarray}
&I^D_{xy}(0,-1)=-\frac{12g\gamma k_0^3}{l}\int_0^{\infty}\frac{q^2dq}{\left[(q^2-k_0^2(1-g))^2+k_0^2/l^2\right]\left[(q^2-k_0^2(1+g))^2+k_0^2/l^2\right]}+\nonumber\\
&+\frac{24g\gamma k_0^3}{l}\int_0^{\infty}\frac{q^2dq}{\left[(q^2-k_0^2(1-g))^2+4k_0^2/l^2\right]\left[(q^2-k_0^2(1+g))^2+4k_0^2/l^2\right]}.
\label{idfin}
\end{eqnarray}
Taking in Eq.(\ref{idfin}) the remaining integrals over $q$ in the limit $k_0l\gg 1$ we finally obtain
\begin{equation}
I_{xy}^D(0,-1)=-\frac{9\pi\gamma l gk_0l}{(2g^2k_0^2l^2+1)(4g^2k_0^2l^2+1)}.
\label{idfinal}
\end{equation}
In the final step using Eq.(\ref{angten}) and dividing $I_{xy}^{D}(0,-1)$ by backscattered intensity at $g=0$, $I_{yy}(0,-1)=3\pi\gamma l$ \cite{migc86}, we arrive at $\tan\Theta_b^D$ of Eq. (\ref{diffcross}) in the limit $gk_0l\ll 1$.

\section[\hspace{2.4cm} Maximally crossed diagrams contribution to the Kerr effect]
{Maximally crossed diagrams contribution to the Kerr effect}
In this section we derive the maximally crossed diagrams contribution to the Kerr effect. i.e., $\tan\Theta_b^C=\frac{ig}{4}$ of Eq. (\ref{diffcross}). To this end, let us represent the maximally crossed diagrams contribution, see Fig.2, to the intensity tensor in the form
\begin{eqnarray}
&I_{ij}^C(\vec R,\vec s)=2\int_0^{\infty}q^2dq\int_{\vec r, \vec r_1, \vec r_2,\vec \rho_1..\vec \rho_4}e^{-iq\vec s\vec r}G^R_{im}(\vec R+\vec r/2,\vec \rho_1)G_{nj}^A(\rho_2,\vec R-\vec r/2) \nonumber \\
&P_{mnhs}^C(\vec\rho_1,\vec\rho_2,\vec\rho_3,\vec\rho_4)G^R_{hl}(\vec\rho_3,\vec r_1)G^A_{fs}(\vec r_2,\vec\rho_4)j_l(\vec r_1)j_f^*(\vec r_2).
\label{max}
\end{eqnarray}
It follows from the symmetry, see Fig.2, that
\begin{equation}
P^C_{mnhs}(\vec\rho_1,\vec\rho_2,\vec\rho_3,\vec\rho_4)=P_{mshn}(\vec\rho_1,\vec\rho_4,\vec\rho_3,\vec\rho_2).
\label{symm}
\end{equation}
Note that throughout the paper we consider the maximally crossed and diffusion propagators $P^C, P$ at $g=0$. The reason is that in the weak scattering limit $k_0l\gg 1$ the main contribution gives the pole term which is unaffected by Faraday rotation or optical activity \cite{MacJon88}.
Using Eqs.(\ref{difmod}) and (\ref{symm}), one has from Eq.(\ref{max})
\begin{equation}
I_{ij}^C(\vec R,\vec s)=2\int_0^{\infty}q^2dq\int d\vec re^{-iq\vec s\vec r}\int d\vec r_1 d\vec r_2 G^R_{is}(\vec R+\vec r/2,\vec r_1)G_{hj}^A(\vec r_2,\vec R-\vec r/2) P(\vec r_1,\vec r_2)\bar E_h(\vec r_2)\bar E_s^*(\vec r_1).
\label{max2}
\end{equation}
Going over to the two and three dimensional Fourier transforms in Eq.(\ref{max2}) and using Eq.(\ref{chpart}), we have
\begin{eqnarray}
&I_{xy}^C(0,\vec s)=4\int_0^{\infty}q^2dq\int_{-\infty}^{+\infty}\frac{dp_z}{2\pi}G_{xy}^R(q\vec s_{\rho},p_z)G^A_{yy}(-q\vec s_{\rho},p_z-2qs_z)\nonumber \\
&\int_0^{\infty}dz_1dz_2P(|q\vec s_{\rho}|,z_1,z_2)\exp\left[-iz_1(p_z+k_0)+iz_2(k_0-p_z+2qs_z)-\frac{z_1+z_2}{2l}\right].
\label{maxfur}
\end{eqnarray}
We remind that $\vec s_{\rho}$ is the projection of the unit vector $\vec s$ on the $xy$ plane. In the backscattered direction $\vec s_{\rho}=0$ and $s_z=-1$. Integrating the equation (\ref{maxfur}) consequentially over $z_1$ and $z_2$, we obtain
\begin{equation}
I_{xy}^C(0,-1)=4\int_0^{\infty}q^2dq\int_{-\infty}^{+\infty}\frac{dp_z}{2\pi}G_{xy}^R(p_z)G^A_{yy}(p_z+2q)F(p_z),
\label{aftint}
\end{equation}
where
\begin{equation}
F(p_z)=\frac{3\gamma i}{l^2(p_z+k_0-i/2l)(k_0-p_z+2qs_z+i/2l)(-2p_z+2qs_z+i/l)}.
\label{fpz}
\end{equation}
Substituting Eqs. (\ref{avgreen}) and (\ref{fpz}) into Eq.(\ref{aftint}), one has
\begin{equation}
I_{xy}^C(0,-1)=I_1(g)-I_1(-g)+I_2(g)-I_2(g),
\label{igea}
\end{equation}
where
\begin{eqnarray}
&I_1(g)=-\frac{3\gamma}{2l^2}\int_0^{\infty}q^2dq\int\frac{dp_z}{2\pi}\frac{1}{(p_z+k_0-i/2l)(p_z-k_0+2q-i/2l)(p_z+2q-i/2l)(p_z-\sqrt{(1+g)k_0^2+ik_0/l})}\times\nonumber\\
&\times\frac{1}{(p_z+\sqrt{(1+g)k_0^2+ik_0/l})(p_z+2q-\sqrt{(1-g)k_0^2-ik_0/l})(p_z+2q+\sqrt{(1-g)k_0^2-ik_0/l})},
\label{1ige}
\end{eqnarray}
and
\begin{eqnarray}
&I_2(g)=-\frac{3\gamma}{2l^2}\int_0^{\infty}q^2dq\int\frac{dp_z}{2\pi}\frac{1}{(p_z+k_0-i/2l)(p_z-k_0+2q-i/2l)(p_z+2q-i/2l)(p_z-\sqrt{(1+g)k_0^2+ik_0/l})}\times \nonumber\\
&\times\frac{1}{(p_z+\sqrt{(1+g)k_0^2+ik_0/l})(p_z+2q-\sqrt{(1+g)k_0^2-ik_0/l})(p_z+2q+\sqrt{(1+g)k_0^2-ik_0/l})}.
\label{2ige}
\end{eqnarray}
The square roots from the complex numbers  are implied in the sense of arithmetic root. In the next step we expand these
roots on $g$ and $1/k_0l$ up to the linear terms
\begin{eqnarray}
&I_1(g)=-\frac{3\gamma}{2l^2}\int_0^{\infty}q^2 dq\int\frac{dp_z}{2\pi}\frac{1}{(p_z+k_0-i/2l)(p_z-k_o+2q-i/2l)(p_z+2q-i/2l)(p_z-k_0-gk_0/2-i/2l)}\times\nonumber\\
&\times\frac{1}{(p_z+k_0+gk_0/2+i/2l)(p_z+2q-k_0+gk_0/2+i/2l)(p_z+2q+k_0-gk_0/2-i/2l)},
\label{exp1ige}
\end{eqnarray}
and
\begin{eqnarray}
&I_2(g)=-\frac{3\gamma}{2l^2}\int_0^{\infty}q^2 dq\int\frac{dp_z}{2\pi}\frac{1}{(p_z+k_0-i/2l)(p_z-k_o+2q-i/2l)(p_z+2q-i/2l)(p_z-k_0-gk_0/2-i/2l)}\times\nonumber\\
&\times\frac{1}{(p_z+k_0+gk_0/2+i/2l)(p_z+2q-k_0-gk_0/2+i/2l)(p_z+2q+k_0+gk_0/2-i/2l)}.
\label{exp2ige}
\end{eqnarray}
We calculate the integral over $p_z$ closing the integral contour in the bottom half of a complex plane. The contribution for $I_1(g)$ comes from the residues of the following two poles
\begin{equation}
 (i). p_z=-k_0-gk_0/2-i/2l\quad (ii). p_z=-2q+k_0+gk_0/2-i/2l.
\label{polesi1ge}
\end{equation}
The result reads
\begin{equation}
I_1(g)=\frac{3i\gamma}{8k_0^2l^2(gk_0+2i/l)}\int_0^{\infty}dq\frac{q+k_0}{(q-k_0+gk_0/4+i/2l)(q-k_0-gk_0/4-i/2l)}.
\label{i1geintq}
\end{equation}
Eq.(\ref{i1geintq}) was derived taking into account the fact, that the main contribution to the integral on $g$ gives the values that are close to $k_0$. Calculating the integral  Eq.(\ref{i1geintq}) in the pole approximation which is justified provided that $k_0l\gg 1$, one has
\begin{equation}
I_1(g)=-\frac{3\pi \gamma}{2k_0^2l^2}\frac{k_0}{(gk_0+2i/l)^2}.
\label{i1gefin}
\end{equation}
Correspondingly
\begin{equation}
I_1(g)-I_1(-g)=\frac{12i\pi \gamma l g}{(4+g^2k_0^2l^2)^2}.
\label{i1gemin}
\end{equation}
In analogous manner one can show that $I_2(g)-I_2(-g)=0$. Therefore using Eqs.(\ref{ige}) and (\ref{i1gemin}), one finds
\begin{equation}
I_{xy}^C(0,-1)=\frac{12i\pi \gamma l g}{(4+g^2k_0^2l^2)^2}.
\label{maxcontr}
\end{equation}
Similarly to what was done in Appendix A, where the $\tan\Theta_b^D$ was calculated, one readily sees that dividing $I_{xy}^{C}(0,-1)$ by backscattered intensity at $g=0$, $I_{yy}(0,-1)=3\pi\gamma l$ \cite{migc86} yields $\tan\Theta_b^C$ of Eq. (\ref{diffcross}) in the limit $gk_0l\ll 1$.






\begin{thebibliography}{99}

\bibitem{and} P. W. Anderson, Phys. Rev. {\bf 109}, 1493 (1958).
\bibitem{Neuw99} M.C.W. van Rosum and Th.M. Nieuwenhuizen, Rev. of Mod. Physics {\bf 71},313,(1999).
\bibitem{ki84} Y.Kuga and A.Ishimaru, J.Opt.Soc.Am. A, {\bf 1}, 831,(1984).
\bibitem{al85} M.P. VanAlbada and A.Lagendijk, Phys. Rev. Lett. {\bf 55}, 2692,(1985).
\bibitem{wm85}P.E.Wolf and G.Maret, Phys. Rev. Lett. {\bf 55}, 2696,(1985).
\bibitem{eta86} M. Kaveh, M. Rosenbluh, I. Edrei, and I. Freund, Phys. Rev. Lett. {\bf 57}, 2049,(1986).
\bibitem{gl97} Zh. S. Gevorkian, and Yu. E. Lozovik, Physica Scripta. {\bf 56}, 208, (1997)
 \bibitem{br12}Matteo Burresi,Vivekananthan Radhalakshmi, Romolo Savo, Jacopo Bertolotti, Kevin Vynck and Diedrik S.Wiersma
{\bf 108},110604,(2012).
\bibitem{migc86} Michael J. Stephen and Gabriel Cwilich, Phys. Rev. B {\bf 34},7564,(1986).
\bibitem{MacJon88}F.C. MacKintosh and Sajeev John, Phys. Rev. B {\bf 37}, 1884, (1988).
 \bibitem{pr03} E.Prati, J. of Electr. Wav. and Appl.{\bf 17, 8}, 1177,(2003).
 \bibitem{lm92} Malcolm Longair, High Energy Astrophysics, Cambridge University Press,(1992).
\bibitem{tktifh06} S.Tomita, T. Kato, S.Tsunashima, S. Iwata, M. Fujii,S. Hayashi, Phys. Rev. Lett. {\bf 96}, 167402,(2006).
\bibitem{fbkui08} R.Fujikawa, A.V.Baryshev, J.Kim, H. Uchida, M. Inoue, J. Appl. Phys. {\bf 103},07D301,(2008).
\bibitem{umfbi09} H.Uchida, Y.Masuda, R.Fujikawa, A.V.Baryshev, M.Inoue, Journal of Mag. and Magnetic Materials {\bf 321},843,(2009).
\bibitem{uch11} H. Uchida, Y. Mazutani, Y. Nakai, A.A. Fedyanin, and M. Inoue, J. Phys. D: Appl. Phys. {\bf 44} 064014 (2011)
\bibitem{tlkrm11} S. Tkachuk, G. Lang, C. Kraft, O. Rabin and I. Mayergoyz, Journal of Applied Physics, {\bf 109}, 07B717,(2011).
\bibitem{dwbwc11} Raj Kumar Dani, Hongwang Wang, Stefan H.Bossman, Gary Wysh and Viktor Chikan, The Journal of Chem.Phys., {\bf 135}, 224502, (2011).
\bibitem{nature11} Iris Grassee, Julien Levallois, Andrew L. Walker, Marus Ostler, Aaron Bostwick, Eli Rotenberg, Thomas Seyller, dirk van der Maret and Alexey Kuzmenko, Nature Physics, {\bf 7}, 48,(2011).
 \bibitem{choy99}  Tuck C.Choy , Effective Medium Theory. Oxford: Clarendon Press (1999).
\bibitem{And85} P. W.Anderson, Philosophical Magazine B, {\bf 52}, 505, (1985).
\bibitem{Gev98}  Zh. S. Gevorkian, Phys. Rev. E  {\bf 57}, 2338 (1998).
\bibitem{ln65} J.S.Langer and T.Neal, Phys. Rev. Lett. {\bf 16}, 984,(1965).
\bibitem{gas}V. Gasparian, M. Ortu{\~n}o, J. Ruiz, E. Cuevas, Phys. Rev. Lett. {\bf 75}, 2312 (1995).
\bibitem{gas1} V. Gasparian, T. Christen, and M. B{\"u}ttiker, Phys. Rev. A {\bf 54}, 4022 (1996).
\bibitem{verdet}Aloke Jain,Jayant Kumar,Fumin Zhou, Lian Li,Sukant Tripathy, Am.J.Phys,{\bf 67}(8),714,(1999).
\bibitem{john85} Sajeev John, Phys. Rev. B, {\bf 31}, 304,(1985).
\bibitem{asb86}  K. Arya, Z.B. Su and Joseph L. Birman, Phys. Rev. Lett., {\bf 57},2725,(1986).
\bibitem{jocher72} P.B.Johnson and R.W.Chersty, Phys. Rev. B, {\bf 6}, 4370,(1972).
\end{thebibliography}
\end{document}